\documentstyle[eqsecnum,prd,aps,epsf]{revtex}

\newcommand{\beq}{\begin{equation}}
\newcommand{\eeq}{\end{equation}}
\newcommand{\beqn}{\begin{eqnarray}}
\newcommand{\eeqn}{\end{eqnarray}}
\newcommand{\pa}{\partial}

\draft
\begin{document}

\twocolumn[\hsize\textwidth\columnwidth\hsize\csname
@twocolumnfalse\endcsname

%======================================%
%<<<<<<<<<<<< TITLE PAGE >>>>>>>>>>>>>>%
%======================================%
%\thispagestyle{empty}

\begin{center}
{\large\bf{Black holes in the brane world 
: Time symmetric initial data }}
~\\
~\\
Tetsuya Shiromizu$^{1,2,3}$ and Masaru Shibata$^{4,5}$ \\
~\\
$^1${\em 
MPI f\"ur Gravitationsphysik, Albert-Einstein Institut, D-14476 Golm, Germany
}\\
$^2${\em 
Department of Physics, The University of Tokyo, Tokyo 113-0033, Japan}\\
$^3${\em Research Center for the Early Universe(RESCEU), 
The University of Tokyo, Tokyo 113-0033, Japan}\\
$^4${\em Department of Physics, University of Illinois at Urbana-Champaign, 
Urbana, IL 61801, USA}\\
$^5${\em Department of Earth and Space Science, 
Osaka University, Toyonaka 560-0043, Japan}
\end{center}

%\date{\today}

%\maketitle

\begin{abstract}
We numerically construct time-symmetric initial data sets of 
a black hole in the Randall-Sundrum brane world model, assuming that 
the black hole is spherical on the brane. We find that 
the apparent horizon is cigar-shaped in the 5D spacetime. 
\end{abstract}
\vskip2pc]

\vskip 1cm 

\section{Introduction}

Motivated by Horava-Witten model \cite{Witten}, 
the so called brane world model 
has been actively investigated \cite{old}. 
Among several models, 
a simple, but very attractive model was recently proposed by 
Randall and Sundrum \cite{RS1,RS2}. 
According to their scenario, we are 
living in a 4D domain wall in 5D 
bulk spacetime. The noteworthy features of 
their model are that in the linearized theory,  
the conventional gravity can be recovered on the brane 
\cite{RS1,RS2,Tama,Misao,Lisa} and that 
a homogeneous, isotropic universe can be simply described if we consider 
a 4D domain wall moving in the 5D Schwarzschild-anti 
de Sitter spacetime \cite{cosmos}. 

One of the most non-linear objects in the theory of gravity 
is a black hole, which should be also investigated to understand 
the nature of the models in strong fields. However, 
because of the complexity of the 
equations, any realistic, exact solutions for black holes have not been 
discovered in the brane world model, 
even with help of numerical computation so far. 
We only know that the effective 4D 
gravitational equation on the brane is different from 
the Einstein equation \cite{Tess} (see Appendix A), so that 
the static solution for a non-rotating black hole should not be identical with 
the 4D Schwarzschild solution. Indeed, a 
linear perturbation analysis \cite{Tama,Lisa} shows that a solution 
of gravitational field outside 
self-gravitating bodies on the brane 
is slightly different from the 4D Schwarzschild solution. 
Chamblin et al. \cite{Andrew} conjecture that the topology of 
black hole event horizons would be spherical with the 
cigar-shaped surface in the 5D spacetime. 
However, nothing has been clarified substantially.

In this paper, as a first step toward self-consistent 
studies for black holes in the 
brane world, we numerically compute 
a black hole space using a time symmetric initial value formulation; 
namely we solve the 5D Einstein equation only on a spacelike 
4D hypersurface. Thus, 
the black hole obtained here is not static nor the exact 
solution for the 5D Einstein equation, implying that 
we cannot identify the event horizon. However, we can investigate 
the property of the horizon determining the apparent horizon which 
could give us an insight on the black hole in the brane world. 
We focus on the Randall-Sundrum's second model \cite{RS2}, 
and assume that the black hole is spherical on the brane, 
but the shape of the horizon is non-trivial in the bulk. 
We will determine the apparent horizon on the brane 
and show that the black hole is cigar-shaped as conjectured in 
\cite{Andrew}.

\section{formulation and results}
%\section{Initial Data of Brane-World Black Holes}

%\subsection{Initial data set}

We consider time symmetric, spacelike hypersurfaces, $\Sigma_t$, 
in the brane world model assuming the vanishing extrinsic 
curvature; i.e., 
%================================%
%
\begin{eqnarray}
H_{\mu\nu} \equiv 
(\delta_\mu^\alpha+t_\mu t^\alpha){}^{(4)} \nabla_\alpha t_\nu=0,
\end{eqnarray}
%
%================================%
where $t^\mu$ is the unit normal timelike vector 
to $\Sigma_t$ and ${}^{(4)} \nabla_\alpha$ is  
the covariant derivative with respect to the 4D metric on $\Sigma_t$.
In this case, the momentum constraint is satisfied trivially, and 
the equation of the Hamiltonian constraint becomes 
%================================%
%
\begin{eqnarray}
{}^{(4)}R=16\pi G_5(\Lambda + {}^{(5)}T_{\mu\nu}t^{\mu}t^{\nu}), 
\end{eqnarray}
%
%================================%
where ${}^{(4)}R$ is the Ricci scalar on $\Sigma_t$, and 
$G_5(=\kappa_5^2/8\pi)$, $\Lambda$ and ${}^{(5)}T_{\mu\nu}$ denote 
the gravitational constant, negative cosmological constant, 
and energy-momentum tensor in 5D spacetime [cf., Eq. (\ref{eq:energy})]. 
We choose the line element on $\Sigma_t$ 
in the form  
%================================%
%
\beq
dl^2=
\frac{1}{z^2}\Bigl[\ell^2 dz^2+\psi^4 (dr^2 + r^2d\Omega) \Bigr], 
\label{eq:metric}
\eeq
%
%================================%
where $\ell=\sqrt{-\kappa_5^2 \Lambda /6}$, 
$z~(\geq 1)$ denotes the coordinate orthogonal to
the brane and $r~(\geq 0)$ is the radial coordinate on the brane. 
We assume that the brane is located at $z=1$. 
Note that we simply choose this line element for convenience of 
the analysis. In this paper, we focus on a 
black hole which is spherical on the brane, i.e., $\psi=\psi(r, z)$. 
Then, the explicit form of the 
Hamiltonian constraint in the bulk (for $z > 1$) is written in the form 
%================================%
%
\beqn
&& \psi''+\frac{2}{r}\psi'+\frac{3}{2\ell^2}\Bigl[ 
\Bigl( \partial_z^2 \psi -\frac{3}{z}\partial_z \psi \Bigr)\psi^4
+3 (\partial_z \psi)^2\psi^3\Bigr] \nonumber \\
&&~~~~~~~~~~~~~=-{\kappa_5^2 \over 4}
{}^{(5)}\tau_{\mu\nu}t^{\mu}t^{\nu}. \label{ham}
\eeqn
%
%================================%
where $'=\partial/\partial r$, and  
${}^{(5)}\tau_{\mu\nu}$ is the energy-momentum tensor 
in the bulk, which is introduced for numerical convenience.

Equation (\ref{ham}) is an elliptic type equation and should be 
solved imposing boundary conditions at $z=1$, $z \gg 1$, $r=0$, and 
$r \gg \ell$. The boundary condition at $z=1$ is derived from Israel's 
junction condition \cite{Israel} as (see Appendix A for the derivation)
%================================%
%
\begin{eqnarray}
\partial_z \psi|_{z=1}=0.\label{eq:bound}
\end{eqnarray}
%
%================================%
The boundary conditions at $z \gg 1$ and $r \gg \ell$ are obtained 
from the linear perturbation analysis (see Appendix B). 
For $r \gg \ell$ and $r > \ell z$, it becomes 
%================================%
%
\begin{eqnarray}
\psi \simeq 1+\frac{MG_4}{2r}\Bigl[1+ 
\frac{1}{2}\Bigl( \frac{R}{r}\Bigr)^2
%+ \Bigl\lbrace \frac{1}{2}\Bigl(\frac{5}{2}+3{\rm ln}(R/2r) \Bigr)-
%\frac{3}{8}[4(z-1)^2+4(z-1)^3+(z-1)^4] \Bigr\rbrace  \Bigl( \frac{R}{r} 
%\Bigr)^4 
+O\Bigl((\ell/r)^4\Bigr) \Bigr],
\end{eqnarray}
%
%================================%
where $G_4=G_5/\ell$, $M$ is the gravitational mass 
on the brane, and $R = (2/3)^{1/2} \ell$. For $z \gg 1$, 
%================================%
%
\begin{eqnarray}
\psi  \simeq 1+\frac{3}{4} \frac{G_4 M}{R z}
\Bigl(1+\frac{r^2}{z^2R^2} \Bigr)^{-3/2}. 
\end{eqnarray}
%
%================================%

To determine the existence of a black hole, 
we search for the apparent horizon.  Here, we 
determine two horizons \cite{Exp}. One is 
defined to be the spherical two-surface on the brane on which 
the expansion of the null geodesic 
congruence confined on the brane is zero\cite{Masaru2}, i.e., 
%================================%
%
\begin{eqnarray}
\theta_3=\frac{2}{\psi^3}\Bigl(2\psi'+\frac{1}{r} \psi \Bigr)=0.
\end{eqnarray}
%
%================================%
The other is the apparent horizon in full 4D space, 
which is defined with respect to the null geodesic congruence 
in full 5D spacetime and satisfies\cite{Masaru2}  
%================================%
%
\begin{eqnarray}
\theta_4={}^{(4)}\nabla_i s^i=0, 
\end{eqnarray}
%
%================================%
where $s^i$ is a unit normal to the surface of the apparent horizon.
Explicit equation for determining this apparent horizon is 
shown in Appendix C. 

The procedure of numerical analysis is 
as follows. First, we artificially put the matter 
of $\rho_h \equiv {}^{(5)}\tau_{\mu\nu}t^{\mu}t^{\nu}>0$ in the bulk. 
This method is employed because we do not have to consider the inner 
boundary condition of black holes with this treatment. 
As long as $\rho_h$ is confined around the brane and inside 
the horizon, it does not significantly 
affect the geometry outside the horizon. 
Then, we solve Eq. (\ref{ham}), and try to find the apparent 
horizon both on the brane and in the bulk. When the distribution of 
$\rho_h$ is sufficiently compact, the apparent horizons exist. 
It should be noted that two horizons do not coincidently appear. 
In some cases, the apparent horizon on the brane exists 
although that in the bulk does not. 

\begin{figure}[t]
\vspace*{-7mm}
\begin{center}
\epsfxsize=2.2in
~\epsffile{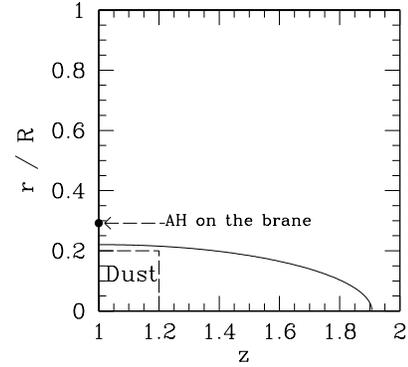}
\end{center}
\vspace*{-8mm}
\caption{Location of the apparent horizons on the brane 
(filled circle) and in the 4D space (solid line). 
Artificial matter is confined in the region shown by the dashed line. 
}
\end{figure}

\begin{figure}[t]
\vspace*{-9mm}
\begin{center}
\epsfxsize=2.2in
~\epsffile{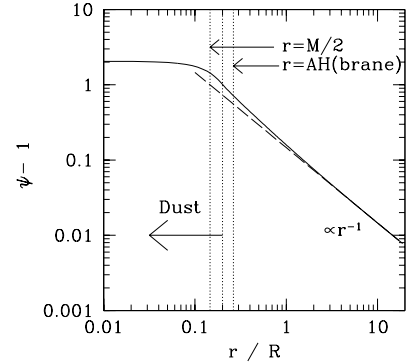}
\end{center}
\vspace*{-8mm}
\caption{Profile of $\psi-1$ on the brane (solid line). 
Location of the apparent horizon on the brane is 
shown. The dashed line denotes $\psi-1=M/2r$}
\end{figure}

Here, we show one example of numerical results. We set $G_4=1$. 
In this example, 
an artificial matter is put for $0 \leq r \leq 0.2 R $ and 
$1 \leq z \leq 1.2$. Equation (\ref{ham}) is solved using a 
uniform grid with grid size $1200 \times 1200$ for $r$ and $z$ 
directions, which covers a domain with $0\leq r/R \leq 17.1$ and 
$1 \leq z \leq 18.1$. In this case, 
the gravitational mass on the brane is 
$M\simeq 0.29R$, and both apparent horizons on the brane and 
in the bulk exist. We note that the results 
are essentially the same for $0.25 \leq M/R \leq 0.5$. 
In Fig. 1, we show the location of apparent horizons in the 
bulk and on the brane. The apparent horizon in the bulk is apparently 
cigar-shaped. Due to this cigar-shape  
the circumferential radius of the apparent 
horizon is different depending on the 
choice of the circumference in the bulk. 
In Fig. 2, we show that the profile of $\psi - 1$ on the brane. 
For $r \gg R$, $\psi-1$ behaves as $M/2r$, implying that 
the solution approximately agrees with that in the 4D Einstein gravity, 
i.e., the bulk effect is small. 
However, the existence of the bulk is significant 
for $r \sim R$ as expected. 
Indeed, $\psi-1$ deviates from $M/2r$ with decreasing 
$r$. This effect is in particular important for the location and 
area of the apparent horizon on the brane: In the case of 4D gravity
without bulk, the apparent horizon is located at $r_{\rm AH}=M/2$ with 
the area $A_{\rm AH}=16\pi M^2$. 
However, in the brane world model, they take different values in general.  
(In this example, $r_{\rm AH} \simeq 0.9M$ and 
$A_{\rm AH} \simeq 88.6M^2$,
and the coefficients converge to well-know 4D values 
(0.5 and $16\pi$) with increasing $M$, implying that the 
effect of the existence of the bulk becomes less important.)

\section{Summary}

We numerically computed time symmetric initial 
data sets of a black hole in the brane world model, 
assuming that the black hole is spherical on the brane. 
As has been expected, the black hole (apparent horizon) is 
cigar-shaped in the bulk \cite{Andrew}. 

We remind that we only present 
time symmetric initial data of a black hole space. 
This implies that the black hole is not static and will evolve 
to other state with time evolution. The quantitative 
features of the final fate could be different from the present 
result. Self-consistent analysis for static black holes 
should be carried out for future 
to obtain a definite answer with regard to black holes in 
the brane world. However, we believe that the present result 
provides us a guideline for such future works. 

\begin{acknowledgments}

We thank B. Carter, N. Dadhich, D. Langlois, 
R. Maartens, K. Maeda, M. Sasaki and T. Tanaka for discussions. 
TS is grateful to thank Relativity Group for their hospitality at Golm 
near by Potsdam. 
MS gratefully acknowledges support by JSPS and the  
hospitality at the Department of Physics of University of Illinois 
at Urbana-Champaign.

\end{acknowledgments}

\appendix

\section{The essence of the brane world}

We briefly review the covariant formalism of the brane world\cite{Tess}. 
For the matter source of the 5D Einstein equation, 
${}^{(5)}G_{\mu\nu}=\kappa_5^2
({}^{(5)}T_{\mu\nu}-\Lambda {}^{(5)}g_{\mu\nu})$, 
%================================%
we choose the energy-momentum tensor as 
%================================%
%
\begin{eqnarray}
{}^{(5)}T_{\mu\nu}=\delta (\chi)
[-\lambda q_{\mu\nu}+{}^{(4)}T_{\mu\nu}]+{}^{(5)}\tau_{\mu\nu},
\label{eq:energy}
\end{eqnarray}
%
%================================%
where $\chi=\ell \ln z$,  
$\lambda$ is the tension of the brane, $q_{\mu\nu}$ is the 
induced metric on the brane, and ${}^{(4)}T_{\mu\nu}$ is the 
energy momentum tensor on the brane. 
Due to the singular source at $\chi=0$ and the $Z_2$ symmetry, 
we can derive the Israel's junction condition at $\chi=0$ as  
%================================%
%
\begin{eqnarray}
K_{\mu\nu}=-\frac{1}{6} \kappa_5^2 \lambda q_{\mu\nu}
-\frac{1}{2} \kappa_5^2\Bigl({}^{(4)}T_{\mu\nu}-\frac{1}{3}q_{\mu\nu}
{}^{(4)}T_{\sigma}^{~\sigma} \Bigr),
\label{eq:ext}
\end{eqnarray}
%
%================================%
where $K_{\mu\nu}=q_{\mu}^{\sigma}q_{\nu}^{\lambda}D_{\sigma}n_{\lambda}$, 
and $D_{\sigma}$ and $n^{\mu}$ are 
the covariant derivative with respect to $q_{\mu\nu}$ and the unit spacelike 
normal vector to the brane. In the text, we consider the cases in which 
${}^{(4)}T_{\mu\nu}=0$. Using (4+1) formalism, the effective 
4D equation on the brane has the form 
%================================%
%
\begin{eqnarray}
{}^{(4)}G_{\mu\nu}=-\Lambda_4 q_{\mu\nu}-E_{\mu\nu}, \label{eq:Einstein}
\end{eqnarray}
%
%================================%
where ${}^{(4)}G_{\mu\nu}$ is the 4D Einstein tensor on the brane, 
%================================%
%
\begin{eqnarray}
\Lambda_4=\frac{1}{2} \kappa_5^2 
\Bigl(\Lambda+\frac{1}{6} \kappa_5^2 \lambda^2 \Bigr)~~{\rm and}~~
E_{\mu\nu} 
=  {}^{(5)}C_{\mu\rho\nu\sigma}n^\rho n^\sigma, 
%& = & -\bL_n K_{\mu\nu}+K_{\mu\alpha} K^\alpha_\nu
%-\frac{1}{6} \kappa_5^2 \Lambda q_{\mu\nu},
\end{eqnarray}
%
%================================%
where ${}^{(5)}C_{\mu\rho\nu\sigma}$ is 5D Weyl tensor.  
%and $\bL_n$ denotes the Lie derivative along $n^{\mu}$. 
In the above, for simplicity, we set ${}^{(5)}\tau_{\mu\nu}=0$. 
Equation (\ref{eq:Einstein}) implies that 
we can consider $E_{\mu\nu}$ as the effective source term of the 
4D Einstein equation on the brane, and 
as long as $E_{\mu\nu}$ is not vanishing, the 
geometry on the brane is different from that in the 4D gravity 
even in the vacuum case.  Only for very special case such as 
for the black string solution \cite{Andrew,Ruth}, 
$E_{\mu\nu}=0$ holds. 

{}From Eq.~(\ref{eq:Einstein}), we find that 
the Minkowski spacetime is realized 
on the brane when $E_{\mu\nu}=0$ and $\Lambda_4=0$. 
In this paper, we set 
$\Lambda_4=0$ to focus on asymptotically flat brane.  Then, 
the junction condition at $\chi=0$ is rewritten to 
$K_{\mu\nu}=-\frac{1}{\ell}q_{\mu\nu}$. 
In the case when we choose the line element as Eq. (\ref{eq:metric}), the 
junction condition reduces to Eq. (\ref{eq:bound}).

\section{Asymptotic boundary conditions}

To specify the boundary condition at infinities, 
we investigate the linearized equation of Eq. (\ref{ham}):
%================================%
%
\begin{eqnarray}
\varphi''+\frac{2}{r}\varphi'+\frac{1}{R^2}
\Bigl( \partial_z^2 \varphi -\frac{3}{z}\partial_z \varphi \Bigr)=
- {\kappa_5^2 \over 4} \rho_h,
\label{eq:lin}
\end{eqnarray}
%
%================================%
where $\psi = 1+\varphi$ and $\varphi \ll 1$. 
We can obtain the formal solution with aid of the 
Green function $G(x,z;x',z')$ as 
%================================%
%
\begin{eqnarray}
\varphi \simeq -2\pi G_4 \ell \int d^3 x'dz' G(x, z;x',z')\rho_h(x', z').
\label{eq:solution}
\end{eqnarray}
%
%================================%
Assuming that $\rho_h$ is non-zero only in the small region 
around the brane, 
we can derive the relevant Green function as\cite{Tama}
%================================%
%
\beqn
G(x,z;x',z')
=&&-\int \frac{d^3k}{(2\pi)^3}e^{i{\bf k}\cdot ({\bf x}-{\bf x}')}
\nonumber \\
&& ~\times 
\Bigl[ \frac{1}{\ell {\bf k}^2}  
+\int_0^\infty dm \frac{u_m(z)u_m(1)}{{\bf k}^2+m^2} \Bigr] 
\nonumber \\
=&& G_0+G_{\rm KK},
\eeqn
%
%================================%
where $u_m(z)$ is the mode function given from the Bessel functions 
$J_n$ and $N_n$ as 
%================================%
%
\begin{eqnarray}
u_m(z)=&& z^2{\sqrt {\frac{mR^2}{2\ell}}} \nonumber \\   
&& \times 
\frac{J_1(mR)N_2(mRz)-N_1(mR)J_2(mRz) }{{\sqrt {(J_1(mR))^2+(N_1(mR))^2}}},
\end{eqnarray}
%
%================================%
where $R=(2/3)^{1/2}\ell$. 
$G_0$ and $G_{\rm KK}$ are the Green function of zero and KK modes, 
respectively. {}From Eq. (\ref{eq:solution}) we can derive the 
asymptotic boundary conditions shown in the text. 

\section{Apparent horizon in the bulk}

We derive the equation for the apparent horizon in the bulk. 
After we perform the coordinate transformation from $(r, z)$ to 
$(x, \theta)$ as 
$z=1+x|\cos\theta|~{\rm and}~r= \ell x\sin\theta$,
the surface of the apparent horizon is denoted by $x=h(\theta)$. 
Then, the non-zero components of $s_i$ is written as 
%================================%
%
\beq
s_x=C~~~{\rm and}~~~s_{\theta}=-C h_{,\theta},
\eeq
%
%================================%
where $C [\equiv \psi^2 \hat C /(1+x|\cos\theta|)]$ 
is a normalization constant calculated from 
$s^is_i=1$, and $h_{,\theta}=dh/d\theta$.
Then, the equation for $h$ can be written to the 
following ordinary differential equation of second order 
%================================%
%
\beqn
{d^2h \over d\theta^2}=&& {h^2 \over \psi^4 {\hat C}^2}\biggl[
\biggl(4{\pa_x \psi \over \psi}+{3 \over 
h(1+h|\cos\theta|)}+{\pa_x \hat C \over \hat C} \biggr) \nonumber \\
&& \times \biggl(\sin^2\theta + \psi^4\cos^2\theta 
- (1 - \psi^4)\sin\theta\cos\theta {h_{,\theta} \over h}\biggr) 
\nonumber \\
&& +h^{-1}\biggl( 4{\pa_{\theta} \psi \over  \psi}+3 {h \sin\theta \over 
1+h|\cos\theta|}+2\cot\theta + D \biggr) \nonumber \\
&& \times \biggl((1 - \psi^4)\sin\theta\cos\theta 
-(\cos^2\theta + \psi^4\sin^2\theta){h_{,\theta}\over h} \biggr)
\nonumber \\
&& +4\psi^3 \pa_{x}\psi (\cos^2\theta 
+ h^{-1} \sin\theta\cos\theta h_{,\theta})\nonumber \\
&& +h^{-2}(1-\psi^4)\sin\theta
\cos\theta h_{,\theta} \nonumber \\
&&
+h^{-1}(1-\psi^4)\cos (2\theta) -4h^{-1}\sin\theta\cos\theta \psi^3 
\pa_{\theta}\psi  \nonumber \\
&& +\{(1-\psi^4)\sin(2\theta)
- 4\sin^2\theta\psi^3\pa_{\theta}\psi\}{ h_{,\theta} \over h^2}
\biggr],
% \nonumber \\
%= && S(\theta, h, h_{,\theta}, \psi, \pa_{\theta} \psi, \pa_x\psi)
\label{AHEQ}
\eeqn
%
%================================%
where 
%================================%
%
\beqn
D=-\hat C^2 [&&(1-\psi^4)\{1-h^{-2} h_{,\theta}^2\}
\sin\theta\cos\theta \nonumber \\
&& -h^{-1}(1-\psi^4)\cos (2\theta) h_{,\theta} 
\nonumber \\
&& +2\psi^3\pa_{\theta}\psi (\cos\theta + h^{-1}\sin\theta h_{,\theta})^2
].
\eeqn
%
%================================%
Eq. (\ref{AHEQ}) is solved imposing boundary conditions 
at $\theta=0$ and $\pi/2$. In the limit $\theta \rightarrow 0$, 
we impose the following boundary condition, 
%================================%
%
\beq
h=h_0 + h_2 \theta^2 + O(\theta^3),
\eeq
%
%================================%
where $h_2$ is evaluated at $x=h_0$ and $\theta=0$ from the 
following equation;  
%================================%
%
\beq
h_2={h_0^2 \over 6}\biggl[
{8 \pa_x \psi \over \psi} + {3  \over h_0(1 + h_0)} 
+{\pa_x \hat C \over \hat C} 
+{3 \over h_0}(1-\psi^4)\biggr]. \label{H2}
\eeq
%
%================================%
At $\theta=\pi/2$, the boundary condition is imposed as $h_{,\theta}=0$. 

Note that in the limit $\theta \rightarrow \pi/2$ (i.e., on the brane), 
Eq. (\ref{AHEQ}) is written in the form  
%================================%
%
\beq
{d^2 h \over d\theta^2}=h + {\ell^2 h^2 \over \psi^4}\biggl(
{4 \pa_x \psi \over \psi} + {2 \over h}\biggr),
\eeq
%
%================================%
where we use $h_{,\theta}=0$ and the relation 
$\pa_{\theta} \psi = D = \pa_x \hat C =0$. 
Note that the equation which the apparent horizon on the brane 
satisfies is $ 4 \pa_x \psi / \psi + 2 /h=0$ [cf., Eq. (2.8)]. 
Thus, unless $d^2 h/d\theta^2 = h$ at $\theta=\pi/2$, 
the apparent horizon on the brane cannot coincide with 
that in 4D space. Note that the black string solution 
\cite{Andrew,Ruth} exceptionally 
satisfies $d^2 h/d\theta^2 = h$ at $\theta=\pi/2$.


\begin{thebibliography}{22}

\bibitem{Witten}
P.~Horava and E.~Witten, Nucl. Phys. {\bf B460}, 506 (1996); {\it ibid}  
{\bf B475}, 94 (1996). 
\bibitem{old}
For earlier work on this topic, see
V. A. Rubakov and M. E. Shaposhinikov, Phys. Lett. {\bf 152B}, 136 (1983);
M. Visser, Phys. Lett. {\bf 159B}, 22 (1985);
M. Gogberashvili, Mod. Phys. Lett. {\bf A14}, 2025 (1999).
\bibitem{RS1}
L.~Randall and R.~Sundrum, Phys. Rev. Lett. {\bf 83}, 3370 (1999).
\bibitem{RS2}
L.~Randall and R.~Sundrum, Phys. Rev. Lett. {\bf 83},4690 (1999).
\bibitem{Tama}
J.~Garriga and T.~Tanaka, Phys. Rev. Lett. {\bf 84}, 2778 (2000).
\bibitem{Misao}
M.~Sasaki, T.~Shiromizu and K.~Maeda, Phys. Rev. D {\bf 62}, 024008 (2000). 
%hep-th/9912233
\bibitem{Lisa}
S.~B.~Giddings, E.~Katz, and L.~Randall, JHEP {\bf 0003}, 023 (2000). 
\bibitem{cosmos}
A.~Chamblin and H.~S.~Reall, Nucl. Phys. {\bf B562},133 (1999); 
T.~Nihei, Phys. Lett. {\bf B465}, 81 (1999);
N.~Kaloper, Phys. Rev. D {\bf 60}, 123506 (1999);
H.~B.~Kim and H.~D.~Kim, Phys. Rev. D {\bf 61}, 064003 (2000);
P.~Binetruy, C.~Deffayet, U. Ellwanger and D. Langlois, Phys. Lett. 
{\bf B477}, 285 (2000);
P.~Kraus, JHEP {\bf 9912}, 01 (1999);
S.~Mukohyama, Phys. Lett. {\bf B473}, 241 (2000);
D.~Ida, gr-qc/9912002; 
J.~Garriga and M.~Sasaki, Phys. Rev. {\bf D62}, 043523 (2000);  
S.~Mukohyama, T.~Shiromizu and K.~Maeda, Phys. Rev. D {\bf 62}, 024028 (2000);
P. Bowcock, C. Charmousis and R. Gregory, hep-th/0007177. 
\bibitem{Tess}
T.~Shiromizu, K.~Maeda and M.~Sasaki, Phys. Rev. D {\bf 62}, 024012 (2000).
%gr-qc/9910076
\bibitem{Andrew}
A.~Chamblin, S.~W.~Hawking and H.~S.~Reall, Phys. Rev. D {\bf 61}, 065007 
(2000).
\bibitem{Israel}
W.~Israel, Nuovo. Cimento. {\bf 44B}, 1 (1996).
\bibitem{Exp}
We note that in the brane world model, photons 
can propagate only on the brane, so that the apparent 
horizon on the brane is with regard to the photons. 
On the other hand, 
the apparent horizon in full 4D space is with regard to 
gravitons which can propagate in full 5D spacetime. 
\bibitem{Masaru2}
For example, M. Shibata, Phys. Rev. D {\bf 55}, 2002 (1997).
\bibitem{Ruth}
R. Gregory, hep-th/0004101.

\end{thebibliography}
\end{document}